\pdfoutput=1
\documentclass[english,letterpaper,conference,compsocconf]{IEEEtran}
\usepackage[T1]{fontenc}
\usepackage[utf8]{inputenc}
\usepackage{mathtools}
\usepackage{algorithm2e}
\usepackage{amsmath}
\usepackage{amsthm}
\usepackage{amssymb}
\usepackage{graphicx}

\makeatletter

\let\labelindent\relax

\usepackage[inline]{enumitem}

\theoremstyle{plain}
\newtheorem{thm}{\protect\theoremname}
\theoremstyle{definition}
\newtheorem{defn}{\protect\definitionname}
\theoremstyle{remark}
\newtheorem{rem}{\protect\remarkname}
\theoremstyle{plain}
\newtheorem{lem}{\protect\lemmaname}

\IEEEoverridecommandlockouts

\makeatother

\usepackage{babel}
\providecommand{\definitionname}{Definition}
\providecommand{\lemmaname}{Lemma}
\providecommand{\remarkname}{Remark}
\providecommand{\theoremname}{Theorem}

\begin{document}

\title{Dynamic Controllability of Conditional Simple Temporal Networks is
PSPACE-complete}

\author{\IEEEauthorblockN{Massimo Cairo\thanks{This work was supported by the Department of Computer Science, University of Verona, under PhD~grant ``Computational Mathematics and
Biology''.}}
\IEEEauthorblockA{Mathematics Department\\ University of Trento\\ Trento, Italy\\  massimo.cairo@unitn.it}
\and
\IEEEauthorblockN{Romeo Rizzi}
\IEEEauthorblockA{Computer Science Department\\ University of Verona\\ Verona, Italy\\  romeo.rizzi@univr.it}}

\maketitle

\global\long\def\Hist{\mathrm{Hist}}
\global\long\def\emptylabel{\lambda}
\global\long\def\Dom{\mathrm{Dom}}

\begin{abstract}
Even after the proposal of various solution algorithms, the precise
computational complexity of checking whether a Conditional Temporal
Network is Dynamically Controllable had still remained widely open.
This issue gets settled in this paper which provides constructions,
algorithms, and bridging lemmas and arguments to formally prove that:
(1) the problem is PSPACE-hard, and (2) the problem lies in PSPACE.
\end{abstract}

\section{Introduction}

In temporal planning and scheduling, a \emph{Simple Temporal Network}
(STN) \cite{Dechter1991} consists of a set of tasks to be scheduled
on the time line, and a set of constraints of the form $Y-X\leq\delta$,
with $\delta\in\mathbb{R}$, i.e., limiting the difference between the
execution times of tasks $X$ and $Y$. The STN is said to be consistent
if it admits a schedule of its tasks that satisfies all the constraints.
Some variants of the STN model have been proposed in the literature to allow and represent
some forms of \emph{contingency}, that is, the presence of parameters
which are unknown to the planner. For example, a \emph{Conditional
Simple Temporal Network} (CSTN) \cite{Tsamardinos2003a} comprises
also a set of unknown propositional variables, and some of the tasks
and constraints in the network are to be taken into account only for
specific values of these variables.

When contingency is present, such as in CSTNs, the notion of consistency
is replaced by the notion of \emph{controllability}, which comes in
three flavors: weak, strong, and dynamic \cite{Vidal1999,Tsamardinos2003a}.
In all the three variants the question is whether the planner is able to provide
a schedule that satisfies the constraints; the difference is in how
and when the value of the unknown parameters is disclosed to the planner.
In the weak controllability, the parameters are revealed before the
execution of the plan, so the schedule can be decided once the value of all the variables (scenario) has been specified, and the main question is deciding whether a feasible scheduling exists for all possible scenarios.
In the strong controllability,
these values are revealed only after the execution of the schedule, so we
need a single schedule that works for every scenario. In the dynamic
controllability, the unknown parameters are revealed progressively
\emph{during} the execution of the plan, as a consequence of actions
performed by the planner. In the case of CSTNs, each propositional
variable is associated with an observation task, and its value is
revealed precisely when the corresponding observation task is executed.
Here we look for a \emph{dynamic execution strategy}: a schedule of
the tasks that gets dynamically decided depending on the partial scenario
progressively observed, such that, whatever scenario possibly
emerges, all the constraints pertinent to that scenario will be respected.
The dynamic controllability\footnote{The term ``dynamic consistency'' is sometimes used in the context
of CSTNs. We prefer to use ``dynamic controllability'' to emphasize
the active role of the planner and to match the name used in the literature
for other type of temporal networks, such as STNUs.} decision problem for CSTNs (CSTN\nobreakdash-DC) asks to check whether
a given CSTN is dynamically controllable, and is a major algorithmic
problem associated to CSTNs.

It is known that the consistency of STNs can be decided in polynomial
time, by interpreting the network as a weighted graph and applying
the Floyd-Warshall All-Pairs Shortest Path algorithm~\cite{Dechter1991}.
However, the presence of contingency might change drastically the
algorithmic nature of the problem. Especially for the dynamic controllability,
which introduces an alternation of quantifiers $\exists$ (for the
choices of the planner) and $\forall$ (for the revealed parameters),
an increase in complexity is expected~\cite{Vidal1999}. In the
case of \emph{Simple Temporal Networks with Uncertainty} (STNUs),
another variant of STNs with contingency (with which a first controllability
issue was posed), the dynamic controllability had been conjectured
to be PSPACE-complete~\cite{Vidal1999}. Subsequently, it was proven
to actually lie in P~\cite{Morris2001}.

For CSTNs, determining the right complexity class of dynamic controllability
is still a widely open question. Deciding their weak controllability
has been proven to be coNP-complete~\cite{Tsamardinos2003a,Comin2015}
and, since weak controllability can easily be reduced to a special
case of dynamic controllability, then CSTN\nobreakdash-DC is at least
coNP-hard \cite{Comin2015}. A first complete algorithmic solution
had been proposed in \cite{Cimatti2014} by reducing CSTN\nobreakdash-DC
to a time automaton game of high complexity. Later, a complete constraint
propagation algorithm was achieved with much better performances in
practice \cite{Hunsberger2015a}, based on the sound constraint-propagation
rules provided in~\cite{Hunsberger2012,Combi2013,Combi2014} for
the more general setting of Conditional Simple Temporal Networks with
Uncertainty (CSTNU). In \cite{Comin2015}, a worst-case upper bound
is obtained, thanks to an algorithm which requires singly-exponential
time and memory. To the best of our knowledge, no better bounds are
known in the literature.

In this work, we settle this question by sharply improving both the
lower and the upper bound. After providing the background notions
and first basic facts as common to both developments in Section~\ref{sec:background},
a reduction from Quantified 3-SAT to CSTN\nobreakdash-DC is proposed
in Section~\ref{sec:hardness}, which proves the latter to be PSPACE-hard.
Then, in Section~\ref{sec:algorithms}, the first algorithm that
solves CSTN\nobreakdash-DC using only polynomial memory is given,
hence showing that CSTN\nobreakdash-DC lies in PSPACE. Sections~\ref{sec:hardness}
and~\ref{sec:algorithms} can be read independently. Taken together,
their negative and positive results show that CSTN-DC is PSPACE-complete,
i.e., the natural complexity class for the dynamic controllability
issue for CSTNs is PSPACE.

\section{Background}

\label{sec:background}In this section, Conditional Simple Temporal
Networks (CSTNs) and their dynamic controllability are formally defined.
The definitions are taken from~\cite{Hunsberger2015a}.

\subsection{Simple Temporal Networks (STNs)}
\begin{defn}[Temporal variables, tasks, constraints]
Let ${\cal T}$ be a finite set of real-valued \emph{temporal variables}.
Each variable $X\in{\cal T}$ represents the execution time of a task,
also denoted with $X$. In the following, we use the terms temporal
variable and task interchangeably. A\emph{ binary difference constraint}
over ${\cal T}$ is a constraint of the form $Y-X\leq\delta$, for
$X,Y\in{\cal T}$ and $\delta\in\mathbb{R}$. In this paper, \emph{constraint}
always denotes a binary difference constraint. The constraint $Y-X\leq\delta$
can also be expressed, equivalently, as $X-Y\geq-\delta$, $Y\leq X+\delta$
or $X\geq Y-\delta$, as it is more convenient in the context.
\end{defn}

\begin{defn}[Schedule, satisfied constraints]
A \emph{schedule} over ${\cal T}$ is a total assignment $\psi\colon{\cal T}\to\mathbb{R}$
of the temporal variables in ${\cal T}$. We write $[\psi]_{X}$ instead
of $\psi(X)$ to denote the value assigned
by the schedule $\psi$ to the variable $X\in{\cal T}$. A schedule \emph{satisfies} a constraint
$Y-X\leq\delta$ if $[\psi]_{Y}-[\psi]_{X}\leq\delta$.
\end{defn}

\begin{defn}[Simple Temporal Network]
A \emph{Simple Temporal Network} (STN) is a pair $({\cal T},{\cal C})$
where ${\cal T}$ is a set of temporal variables and ${\cal C}$ is
a set of constraints over ${\cal T}$.
\end{defn}

\begin{defn}[Feasible schedule]
A schedule $\psi$ over ${\cal T}$ is \emph{feasible} for $({\cal T},{\cal C})$
if $\psi$ satisfies all the constraints in ${\cal C}$.
\end{defn}

\subsection{Conditional Simple Temporal Networks (CSTNs)}
\begin{defn}[Propositional variables, labels]
Let ${\cal P}$ be a set of propositional (boolean) variables. A
\emph{label} $\ell$ over ${\cal P}$ is a boolean formula $\ell=l_{1}\land\dots\land l_{k}$,
obtained as conjunction of positive or negative literals $l_{i}\in\{p_{i},\neg p_{i}\}$
on distinct variables $p_{i}\in{\cal P}$. The empty label is denoted
with $\lambda$ and always evaluates to true. Let ${\cal P}^{*}$
denote the set of labels over ${\cal P}$ (including $\lambda$).
\end{defn}

\begin{defn}[Scenario, label evaluation]
A \emph{scenario} $s$ over ${\cal P}$ is a total assignment of
the propositional variables $s\colon{\cal P}\to\{0,1\}$ where $0$
means \emph{false} and $1$ means \emph{true}. Let $\Sigma_{{\cal P}}$
denote the set of all the scenarios over ${\cal P}$. We write $s\vDash\ell$
if the label $\ell$ evaluates to true under the interpretation given
by~$s$.
\end{defn}

\begin{defn}[Conditional Simple Temporal Network]
A \emph{Conditional Simple Temporal Network} (CSTN) is a tuple $\Gamma=({\cal T},{\cal P},{\cal C},{\cal L},{\cal OT},{\cal O})$
where 
\begin{itemize}
\item ${\cal T}$ is a finite set of \emph{temporal variables }or\emph{
tasks},
\item ${\cal P}$ is a finite set of \emph{propositional variables}, 
\item ${\cal C}$ is a finite set of \emph{labeled constraints} $(Y-X\leq\delta,\ell)$,
where $Y-X\leq\delta$ is a constraint over ${\cal T}$ and $\ell\in{\cal P}^{*}$
is a label, 
\item ${\cal L}\colon{\cal T}\to{\cal P}^{*}$ if a function that assigns
a label ${\cal L}(X)$ to each task $X\in{\cal T}$, 
\item ${\cal OT}\subseteq{\cal T}$ is the set of \emph{observation tasks}, 
\item ${\cal O}\colon{\cal P}\to{\cal OT}$ is a bijection that associates
each propositional variable $p\in{\cal P}$ to a unique observation
task ${\cal O}(p)$.
\end{itemize}
\end{defn}
A task $X\in{\cal T}$ has to be executed only in those scenarios
$s\in\Sigma_{{\cal P}}$ such that $s\vDash{\cal L}(X)$, and each
constraint $(Y-X\leq\delta,\ell)\in{\cal C}$ has to be satisfied
if $s\vDash\ell$. Since the constraint $Y-X\leq\delta$ only makes
sense if both $X$ and $Y$ get executed, we require the following
well-definedness property.
\begin{defn}[Restriction WD1]
A CSTN satisfies the restriction \emph{WD1} if, for every labeled
constraint $(Y-X\leq\delta,\ell)\in{\cal C}$, we have $\ell\Rightarrow{\cal L}(X)\land{\cal L}(Y)$.
\end{defn}
In the following, \emph{WD1} is incorporated in the definition of
CSTN, i.e., it is assumed that any CSTN satisfies this restriction.
\begin{rem}
Tsamardinos et al.~\cite{Tsamardinos2003a} discussed some supplementary \emph{reasonability
assumptions} that any well-defined CSTN must satisfy. Subsequently,
those conditions have been analyzed and formalized in \cite{Hunsberger2012}
introducing the three restrictions \emph{WD1}, \emph{WD2}, and \emph{WD3}.
The restriction \emph{WD1} has already been discussed. The restrictions
\emph{WD2} and \emph{WD3} relate the labels on tasks and constraints
with the labels on observation tasks. We avoid entering into the fine
details regarding them, and we rather provide both of our results
in their strongest and most general form, as follows. First, the reduction
in our PSPACE-hardness proof constructs only CSTNs that comply with
all three restrictions vacuously, having no labels on the tasks. Second,
neither \emph{WD2} nor \emph{WD3} are required as preconditions for
the applicability of our PSPACE algorithm.\footnote{The reader may observe that, without \emph{WD2} and \emph{WD3}, it
is possible to have the corner case of a network which does not admit
any dynamic execution strategy~\cite{Hunsberger2012}. Such a network
is considered not dynamically controllable since, in particular, it
does not admit any \emph{viable} dynamic execution strategy. No special
handling of this case is needed.} 
\end{rem}

\subsection{Dynamic controllability of CSTNs}
\begin{defn}[Projection]
The \emph{projection} of a CSTN over a scenario $s$ is the STN $\Gamma_{s}=({\cal T}_{s},{\cal C}_{s})$
where: 
\begin{itemize}
\item ${\cal T}_{s}=\{X\in{\cal T}\mid s\vDash{\cal L}(X)\}$ 
\item ${\cal C}_{s}=\{Y-X\leq\delta\mid(Y-X\leq\delta,\ell)\in{\cal C}\text{ and }s\vDash\ell\}$.
\end{itemize}
\end{defn}

\begin{defn}[Execution strategy, viable]
Let $\Psi_{{\cal T}}$ denote the set of schedules $\psi$ over any
subset ${\cal T}'\subseteq{\cal T}$. For a schedule $\psi\in\Psi_{{\cal T}}$
over ${\cal T}'\subseteq{\cal T}$, let $\Dom(\psi)={\cal T}'$. An
\emph{execution strategy} for $\Gamma$ is a function $\sigma\colon\Sigma_{{\cal P}}\to\Psi_{{\cal T}}$
that maps each scenario $s\in\Sigma_{{\cal P}}$ to a schedule $\sigma(s)$
for $\Gamma_{s}$ (i.e.\ $\Dom(\sigma(s))={\cal T}_{s}$). An execution
strategy $\sigma$ is \emph{viable} if, for every scenario $s\in\Sigma_{{\cal P}}$,
the schedule $\sigma(s)$ is feasible for $\Gamma_{s}$.
\end{defn}

\begin{defn}[Partial scenario, history]
A\emph{ partial scenario} over ${\cal P}$ is a partial assignment
$h\colon\Dom(h)\to\{0,1\}$ of the propositional variables, where
$\Dom(h)\subseteq{\cal P}$. Given $\sigma$, a scenario $s$, and
a time point $t\in\mathbb{R}$, the \emph{history} at $t$ in the
scenario $s$ with the strategy $\sigma$ is the partial scenario
$\Hist(t,s,\sigma)$ where $\Dom(\Hist(t,s,\sigma))=\{p\in{\cal P}\mid[\sigma(s)]_{{\cal O}(p)}<t\}$
and $\Hist(t,s,\sigma)(p)=s(p)$ for every $p\in\Dom(\Hist(t,s,\sigma))$.
\end{defn}

\begin{defn}[Dynamic execution strategy]
An execution strategy $\sigma$ is \emph{dynamic} if, for any scenarios
$s,s'\in\Sigma_{{\cal P}}$ and time variable $X\in{\cal T}_{s}$,
letting $t=[\sigma(s)]_{X}$, if $\Hist(t,s,\sigma)=\Hist(t,s',\sigma)$,
then $X\in{\cal T}_{s'}$ and $[\sigma(s')]_{X}=t$.
\end{defn}

\begin{defn}[Dynamic controllability]
A CSTN is \emph{dynamically controllable} if it admits a dynamic
viable execution strategy. The \emph{dynamic controllability} decision
problem (CSTN\nobreakdash-DC) asks to check whether a given CSTN
$\Gamma$ is dynamically controllable or not.
\end{defn}
The following definition and lemma state a useful characterization
of dynamic execution strategies: if two scenarios differ in only one
propositional variable, then a dynamic execution strategy behaves
in the same way in the two scenarios until that propositional variable
is observed. This property has been also proven in~\cite[Theorem~1]{Hunsberger2015a},
and will be used several times in this paper to exploit the fact that
an execution strategy is dynamic.
\begin{defn}
Given a scenario $s\in\Sigma_{{\cal P}}$, a propositional variable
$p\in{\cal P}$ and $v\in\{0,1\}$, let $s[v/p]$ be the scenario
obtained from $s$ by changing the value of the variable $p$ to $v$,
i.e., $s[v/p](p)=v$ and $s[v/p](q)=s(q)$ for $q\in{\cal P}\setminus\{p\}$. \end{defn}
\begin{lem}
\label{lem:dynamic-tech} Let $\sigma$ be a dynamic execution strategy.
Let $s\in\Sigma_{{\cal P}}$, $p\in{\cal P}$, $v\in\{0,1\}$, and
consider the scenario $s'=s[v/p]$. For any $t\leq[\sigma(s)]_{{\cal O}(p)}$,
the following properties hold:
\begin{enumerate}
\item[(a)] $\Hist(t,s,\sigma)=\Hist(t,s',\sigma)$,
\item[(b)] $[\sigma(s)]_{X}\leq t\iff[\sigma(s')]_{X}\leq t$ for every $X\in{\cal T}$,
\item[(c)] $[\sigma(s)]_{X}=t\iff[\sigma(s')]_{X}=t$ for every $X\in{\cal T}$
\end{enumerate}
and in particular $[\sigma(s)]_{{\cal O}(p)}=[\sigma(s')]_{{\cal O}(p)}$.\end{lem}
\begin{IEEEproof}
Observe that we only need to check the properties for those values
of $t$ that appear somewhere in $\sigma$. Hence, we can proceed
by induction. The properties clearly hold for a sufficiently small
$t$ (less than any value in $\sigma$). Take $t<t'\leq[\sigma(s)]_{{\cal O}(p)}$,
such that $[\sigma(s)]_{X},[\sigma(s')]_{X}\notin(t,t')$ for every
$X\in{\cal T}$, and assume that the properties hold for $t$.
Observe
that, for $s''\in\{s,s'\}$, $\Dom(\Hist(t',s'',\sigma))=\Dom(\Hist(t,s'',\sigma))\cup\{q\in{\cal P}\mid[\sigma(s'')]_{{\cal O}(q)}=t\}$ and, since $[\sigma(s)]_{{\cal O}(q)}=t\iff[\sigma(s')]_{{\cal O}(q)}=t$
for $q\in{\cal P}$, then $\Dom(\Hist(t',s,\sigma))=\Dom(\Hist(t',s',\sigma))$.
Moreover, $p\notin\Dom(\Hist(t',s,\sigma))$ since $t<[\sigma(s)]_{{\cal O}(p)}$
by assumption. Hence, $\Hist(t',s,\sigma)=\Hist(t',s',\sigma)$. The
other two properties for $t'$ are a direct consequence of the fact
that $\sigma$ is dynamic, and this concludes the induction.
\end{IEEEproof}

\section{PSPACE-hardness}

\label{sec:hardness}In this section, we prove that CSTN\nobreakdash-DC
is PSPACE-hard by showing a reduction from Quantified $3$-SAT (Q3SAT).

We are given a Q3SAT formula 
$
\Phi=\exists x_{1}\forall y_{1}\cdots\exists x_{n}\forall y_{n}\:\varphi
$
where $\varphi$ is a $3$CNF over the propositional variables $x_{1},y_{1},\ldots,x_{n},y_{n}$.
I.e., 
$
\varphi=\bigwedge_{j=1}^{m}(l_{j,1}\vee l_{j,2}\vee l_{j,3})
$
 and each literal $l_{j,k}$ is either a positive or a negated occurrence
of one of the quantified variables. The formula $\Phi$ can be understood
as a game in which the existential player and the universal player
decide in turn the value of the variables $x_{1},y_{1},x_{2},y_{2},\ldots,x_{n},y_{n}$.
The existential players wins if, when all the variables have been
set, the formula $\varphi$ is satisfied by the chosen values. CSTNs
can be also seen as games, where the planner plays against the nature,
the first by scheduling the tasks, the second by choosing the value
of the propositional variables as soon as they are observed. The planner
wins if, eventually, the schedule he executes is feasible, and the
CSTN is dynamically controllable if the planner has a winning strategy.
This interpretation of both Q3SAT and CSTN-DC as two-player games
underlies our proof of PSPACE-hardness.

We will describe a CSTN $\Gamma_{\Phi}$ which is dynamically controllable
iff $\Phi\equiv\text{true}$, that is, iff the existential player
has a winning strategy for $\Phi$. It will be apparent that $O(\log(n+m))$
internal space suffices in order to construct $\Gamma_{\Phi}$ out
from $\Phi$.

\subsection{Warm-up: the controller can choose some variables}

Before addressing CSTN\nobreakdash-DC, we consider a more general
problem CSTN$^{+}$\nobreakdash-DC. We define a CSTN$^{+}$ to be
a CSTN in which the values of a subset ${\cal P}^{+}\subseteq{\cal P}$
of the propositional variables are actually \emph{decided} by the
controller rather than by the nature, still each $p\in{\cal P}$ gets
determined at the precise execution time of the corresponding disclosure
task ${\cal O}(p)$. To ease our exposition, we first construct a
CSTN$^{+}$ $\Gamma_{\Phi}^{+}$ which is dynamically controllable
iff $\Phi\equiv\text{true}$. This will be a much easier task, but
helps in delivering the general idea of the reduction.

The CSTN$^{+}$ $\Gamma_{\Phi}^{+}$ contains all the variables $x_{i},y_{i}$
as propositional variables, decided and observed respectively in tasks
$X_{i}={\cal O}(x_{i})$ and $Y_{i}={\cal O}(y_{i})$. These tasks
are subject to the unlabeled constraints $Y_{i}\geq X_{i}+1$ ($i=1,\dots,n$)
and $X_{i+1}\geq Y_{i}+1$ ($i=1,\dots,n-1$). These constraints connect
the tasks $X_{1},Y_{1},\dots,X_{n},Y_{n}$ in a chain which enforces
that they are executed in the proper order. Then, we have two tasks
$A$ and $B$ with the following constraints. For every $j=1,\dots,m$,
we have a constraint $B\geq A+1$ with label $\ell_{j}\coloneqq\neg l_{j,1}\land\neg l_{j,2}\land\neg l_{j,3}$,
defined as the negation of the $j$-th clause $(l_{j,1}\vee l_{j,2}\vee l_{j,3})$
of $\varphi$. Finally, there is an unlabeled constraint $A\geq B+1$.

The network $\Gamma_{\Phi}^{+}$ is dynamically controllable iff $\Phi\equiv\text{true}$.
Indeed, if $\Phi\equiv\text{true}$, the controller schedules $X_{i}$
and $Y_{i}$ at time $2i$ and $2i+1$ respectively, and can choose
the propositional value of $x_{i}$ depending on $y_{1},\dots,y_{i-1}$
in accordance to his winning strategy for $\Phi$. Finally, he schedules
$B$ at time $2n+2$ and $A$ at time $2n+3$, and, since every clause
of $\varphi$ is satisfied, none of the constraints $B\geq A+1$ applies
and all the other constraints are fulfilled. Conversely, assume $\Phi\equiv\text{false}$.
It is now nature that owns a winning strategy: since the controller
is anyhow forced to reveal the variables in order, she can choose
each propositional variable $y_{i}$ depending on $x_{1},\ldots,x_{i}$
so that, for at least one $j\in\{1,\ldots,m\}$, the clause $(l_{j,1}\vee l_{j,2}\vee l_{j,3})$
evaluates to false. Hence, the constraints $A\geq B+1$ and $B\geq A+1$
necessarily lead to a conflict, no matter when the events $A$ and
$B$ are scheduled.

\subsection{Reduction for CSTNs}

The above toy reduction with $\Gamma_{\Phi}^{+}$ illustrates well
the general framework, but relies on the strong assumption that the
controller can \emph{choose} the value of some of the propositional
variables. In CSTNs, the controller cannot force the nature to choose
a particular value for a propositional variable. However, he can put
a lot of pressure on her to choose the value he wants. Indeed, we
next describe a network (in the standard framework of CSTNs) that
allows the controller to specify the value he desires for a variable
$x_{i}$, by executing one among two particular actions, one for true
and one for false. The network is built in such a way that, if the
value actually chosen by the nature differs from the prescription
of the planner, then he is able to schedule the rest of the network
easily, satisfying all the remaining constraints. Thanks to this property,
the nature is effectively obliged to choose the variables as specified
by the planner, otherwise she is doomed to lose the match. 

\begin{figure*}
\begin{centering}
\includegraphics{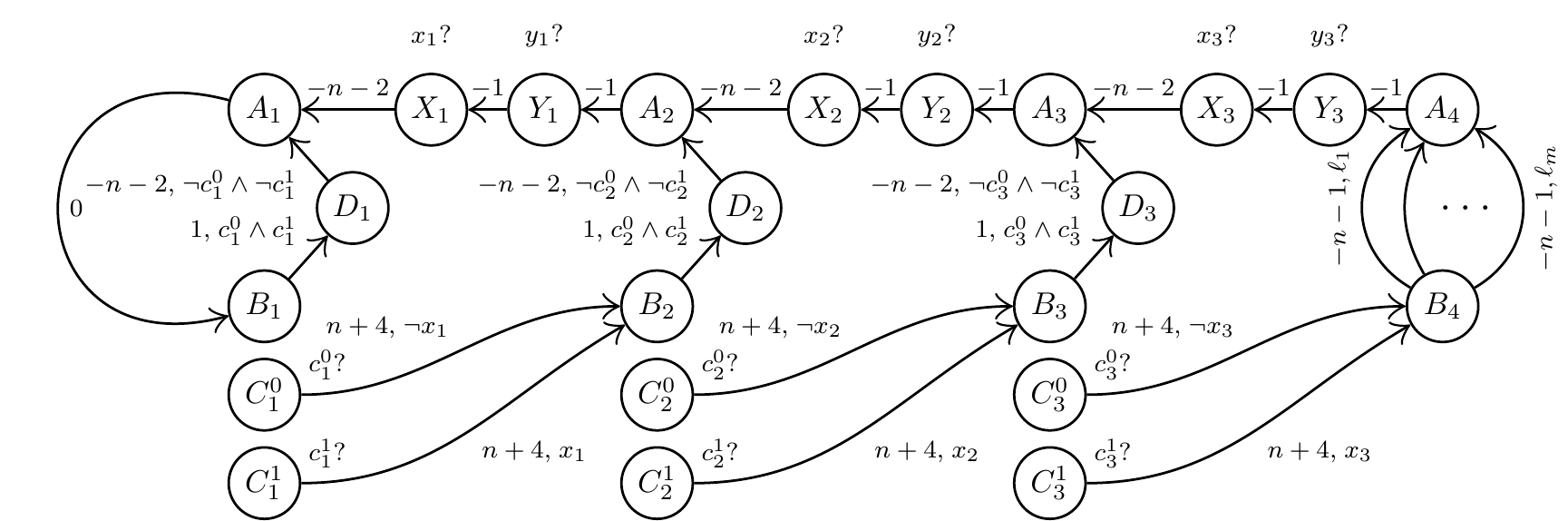}
\par\end{centering}

\caption{\label{fig:reduction}Reduction from Q3SAT to CSTN\protect\nobreakdash-DC
for $n=3$. A Q3SAT formula $\Phi=\exists x_{1}\forall y_{1}\,\exists x_{2}\forall y_{2}\,\exists x_{3}\forall y_{3}\:\bigwedge_{j=1}^{m}(l_{j,1}\vee l_{j,2}\vee l_{j,3})$
is transformed into the network $\Gamma_{\Phi}$ illustrated above,
where nodes denote tasks and edges denote constraints. Specifically,
a directed edge labeled $\delta,\ell$ from a node $N$ to a node
$M$ denotes the labeled constraint $(M\leq N+\delta,\ell)$. The
label $\ell_{j}$ is defined as the negation of the $j$-th clause
of the 3CNF formula, i.e., $\ell_{j}\protect\coloneqq\neg l_{j,1}\land\neg l_{j,2}\land\neg l_{j,3}$.
The empty label $\lambda$ is omitted. A label $q?$ beside a node
$Q$ indicates that $Q={\cal O}(q)$.
}

\end{figure*}

We begin with an informal description of our construction~$\Gamma_{\Phi}$.
Figure~\ref{fig:reduction} shows an example of our construction
for $n=3$, and may help the reader in following the exposition. There
are $n$ gadgets $G_{1},\dots,G_{n}$ connected in series. The $i$-th
gadget $G_{i}$ involves the propositional variables $x_{i},y_{i}$
(which are now normal variables chosen by the nature), and two extra
variables $c_{i}^{1}$ and~$c_{i}^{0}$. The purpose of $G_{i}$
is to let the controller choose the value of $x_{i}$, and then observe
the value of $y_{i}$ chosen by the nature. The nodes of $G_{i}$
are $A_{i}$, $B_{i}$, $C_{i}^{0}={\cal O}(c_{i}^{0})$, $C_{i}^{1}={\cal O}(c_{i}^{1})$,
$D_{i}$, $X_{i}={\cal O}(x_{i})$ and $Y_{i}={\cal O}(y_{i})$. Moreover,
$G_{i}$ connects also to the nodes $A_{i+1}$ and $B_{i+1}$ which,
for $i<n$, belong to the next gadget $G_{i+1}$, while $A_{n+1}$
and $B_{n+1}$ are two extra nodes at the end of the construction.
It is here, between $A_{n+1}$ and $B_{n+1}$, that the $m$ clause
constraints get lied down. For each $j=1,\dots,m$, we put a constraint
$B_{n+1}-A_{n+1}\geq n+1$ with label $\ell_{j}\coloneqq\neg l_{j,1}\land\neg l_{j,2}\land\neg l_{j,3}$,
defined as the negation of the $j$-th clause of $\varphi$, like
in the toy reduction of the previous section. 

Before describing the internals of each gadget, we show how they play
together and we focus only on the tasks $A_{i}$ and $B_{i}$ for
$i=1,\dots,n+1$. Consider the constraint $B_{i}-A_{i}\geq i-1$ for
$1\leq i\leq n+1$, called ``activation constraint''. The gadget
$G_{i}$ is ``activated'' if the $i$-th activation constraint is
satisfied, i.e., if the task $B_{i}$ is executed at most $i-1$ units
of time after $A_{i}$. For the first gadget $G_{1}$, the activation
constraint $B_{1}-A_{1}\leq0$ is explicitly added to the network,
without labels, enforcing the gadget $G_{1}$ to be always activated.
Thanks to the internal structure of the gadgets, the activation constraint
is then propagated from one gadget to the next, as long as the nature
chooses the value of $x_{i}$ according to the prescription of the
controller. If the nature always chooses $x_{i}$ according to the
controller, then all the gadgets are activated and we end up with
the propagated constraint $B_{n+1}-A_{n+1}\leq n$. At this point,
the controller is able to schedule $A_{n+1}$ and $B_{n+1}$ if and
only if all the clauses of $\varphi$ are satisfied, so that the constraints
$B_{n+1}-A_{n+1}\geq n+1$ labeled with $\ell_{i}$ are all void. 

 If instead the nature chooses for any $x_{i}$ the opposite value
to the one prescribed, then the activation constraint on $A_{i}$
and $B_{i}$ is not propagated to $A_{i+1}$ and $B_{i+1}$, the following
gadgets are not activated, and the controller is able to execute all
the other tasks $B_{i'}$, $C_{i'}^{h}$ and $D_{i'}$ for $i'>i$
very far in the future, without violating any constraint.

We now describe the internal mechanism of each gadget. At the heart
of $G_{i}$ there are the two constraints $D_{i}\leq B_{i}+1$ and
$D_{i}\geq A_{i}+(n+2)$, labeled with $c_{i}^{1}\land c_{i}^{0}$
and $\neg c_{i}^{1}\land\neg c_{i}^{0}$ respectively. If the gadget
is activated, then these two constraints cannot be satisfied together
(since $B_{i}-A_{i}\leq i-1\leq n$). Hence, the controller has to
observe either $c_{i}^{1}$ or $c_{i}^{0}$, in order to decide whether
to execute $D_{i}$ early or not (see Lemma~\ref{lem:hard-c0-or-c1}).
Which of the two variables is observed specifies the desired value
of $x_{i}$: so, if the planner wants $x_{i}$ to be true, then he
should execute $C_{i}^{1}$, and if he wants $x_{i}$ to be false,
he should execute $C_{i}^{0}$. A constraint $X_{i}\geq A_{i}+(n+2)$
ensures that the controller can observe $x_{i}$ only after choosing
either $C_{i}^{1}$ or $C_{i}^{0}$. Then, the constraint $Y_{i}\geq X_{i}+1$
allows the controller to observe $y_{i}$ only after $x_{i}$, and
the constraint $A_{i+1}\geq Y_{i}+1$ connects to the next gadget. 

The propagation of the activation constraint $B_{i}-A_{i}\leq i-1$
to the next gadget is achieved by the two constraints $B_{i+1}\leq C_{i}^{1}+(n+4)$
and $B_{i+1}\leq C_{i}^{0}+(n+4)$, labeled with $x_{i}$ and $\neg x_{i}$
respectively. In order for the propagation to take place, the nature
has to choose $x_{i}$ to true if $C_{i}^{1}$ has been executed,
and to false if $C_{i}^{0}$ has been executed (see Lemma~\ref{lem:hard-propagation}).

The full construction is provided for reference in Figure~\ref{fig:construction},
and it is illustrated in Figure~\ref{fig:reduction} for $n=3$.

\begin{figure}[t]
\rule{1\columnwidth}{1pt}
\small

\vspace{1ex}

\mbox{%
\begin{minipage}[t]{0.97\columnwidth}%
\begin{itemize}[label={},itemsep=0.5ex,leftmargin=0ex]
\item  ${\cal T}=\{A_{i},B_{i},C_{i}^{0},C_{i}^{1},D_{i},X_{i},Y_{i}\}_{i=1,\dots,n}\cup\{A_{n+1},B_{n+1}\}$
\item ${\cal P}=\{x_{i},y_{i},c_{i}^{1},c_{i}^{0}\}_{i=1,\dots,n}$,
\item ${\cal L}(N)=\emptylabel$ for every $N\in{\cal T}$,
\item ${\cal C}$ contains the following constraints:

\begin{itemize}[label={},itemsep=0.2ex]
\item $(B_{1}-A_{1}\leq0,\emptylabel)$,
\item for $i=1,\dots,n$:

\begin{itemize}[label={},itemsep=0ex]
\item $(D_{i}\leq B_{i}+1,c_{i}^{0}\land c_{i}^{1})$,
\item $(D_{i}\geq A_{i}+(n+2),\neg c_{i}^{0}\land\neg c_{i}^{1})$,
\item $(X_{i}\geq A_{i}+(n+2),\emptylabel)$,
\item $(Y_{i}\geq X_{i}+1,\emptylabel)$,
\item $(A_{i+1}\geq Y_{i}+1,\emptylabel)$,
\item $(B_{i+1}\leq C_{i}^{0}+(n+4),\neg x_{i})$,
\item $(B_{i+1}\leq C_{i}^{1}+(n+4),x_{i})$,
\end{itemize}
\item for $j=1,\dots,m$:

\begin{itemize}[label={},itemsep=0ex]
\item $(B_{n+1}-A_{n+1}\geq n+1,\neg l_{j,1}\land\neg l_{j,2}\land\neg l_{j,3})$,
\end{itemize}
\end{itemize}
\item ${\cal OT}=\{X_{i},Y_{i},C_{i}^{0},C_{i}^{1}\}_{i=1,\dots,n}$,
\item ${\cal O}(x_{i})=X_{i}$, ${\cal O}(y_{i})=Y_{i}$, ${\cal O}(c_{i}^{0})=C_{i}^{0}$
and ${\cal O}(c_{i}^{1})=C_{i}^{1}$,\\
\hspace*{\fill}for $i=1,\dots,n$.\end{itemize}
\end{minipage}%
}

\vspace{1ex}

\rule{1\columnwidth}{1pt}

\caption{\label{fig:construction}Construction of the CSTN $\Gamma_{\Phi}=({\cal T},{\cal P},{\cal C},{\cal L},{\cal OT},{\cal O})$
for a given Q3SAT formula $\Phi=\exists x_{1}\forall y_{1}\cdots\exists x_{n}\forall y_{n}\:\bigwedge_{j=1}^{m}(l_{j,1}\vee l_{j,2}\vee l_{j,3})$.}

\vspace*{-1em}

\end{figure}

\begin{lem}
\label{lem:easyPSPACE-hard} If $\Phi\equiv\text{true}$ then $\Gamma_{\Phi}$
is dynamically controllable.\end{lem}
\begin{IEEEproof}
Assume $\Phi\equiv\text{true}$. This means that the existential player
holds a winning strategy for $\Phi$. This strategy can be expressed
as a function $f\colon\{0,1\}^{*}\to\{0,1\}$, such that $\varphi$
evaluates to \emph{true} over all truth-assignments $s\colon\{x_{1},y_{1},\ldots,x_{n},y_{n}\}\to\{0,1\}$
in which $s(x_{i})=f(s(y_{1}),\ldots,s(y_{i-1}))$ for every $i=1,\ldots,n$.
Taking $f$ as reference, we provide a viable and dynamic execution
strategy for $\Gamma_{\Phi}$.

Given a scenario $s\colon{\cal P}\to\{0,1\}$, let $h_{i}(s)=f(s(y_{1}),\dots,s(y_{i-1}))$
for $i=1,\dots,n$. Then, define $b(s)$ to be the smallest index
$i\in\{1,\dots,n\}$ such that $s(x_{i})\neq h_{i}(s)$, or $b(s)=n+1$
if no such index $i$ exists. The value $b(s)$ represents the first
index $i$ in which the nature does not follow the prescription of
the controller in choosing the value of $x_{i}$, or $b(s)=n+1$ if
she copies until the end.

The execution strategy $\sigma$ is defined in Figure~\ref{fig:easy-strategy},
where $b'(s)\coloneqq\min\{b(s),n\}$ and $\infty$ denotes a sufficiently
large value, say, $\infty\coloneqq(n+4)(n+2)$. 

\begin{figure}
\small
\rule{1\columnwidth}{1pt}

\vspace{1ex}
$\begin{alignedat}{3} & {}[\sigma(s)]_{A_{i}} &  & =(n+4)\,i\quad &  & \text{for }i=1,\dots,n+1\\
 & [\sigma(s)]_{B_{i}} &  & =(n+4)\,i &  & \text{for }i=1,\dots,b(s)\\
 & [\sigma(s)]_{B_{i}} &  & =\infty &  & \text{for }i=b(s)+1,\dots,n+1\\
 & [\sigma(s)]_{C_{i}^{h}} &  & =(n+4)\,i &  & \text{for }i=1,\dots,b'(s)\text{ and }h=h_{i}(s)\\
 & [\sigma(s)]_{C_{i}^{h}} &  & =\infty &  & \text{for }i=1,\dots,b'(s)\text{ and }h\neq h_{i}(s)\\
 & [\sigma(s)]_{C_{i}^{h}} &  & =\infty &  & \text{for }i=b'(s)+1,\dots,n\text{ and }h=0,1\\
 & [\sigma(s)]_{D_{i}} &  & =\mathrlap{\begin{cases}
(n+4)\,i+1 & \text{if }s(c_{i}^{h_{i}(s)})=1\\
\infty & \text{otherwise }
\end{cases}}\\
 &  &  &  &  & \text{for }i=1,\dots,b'(s)\\
 & [\sigma(s)]_{D_{i}} &  & =\infty &  & \text{for }i=b'(s)+1,\dots,n\\
 & [\sigma(s)]_{X_{i}} &  & =\mathrlap{(n+4)\,i+n+2\qquad\text{for }i=1,\dots,n+1}\\
 & [\sigma(s)]_{Y_{i}} &  & =\mathrlap{(n+4)\,i+n+3\qquad\text{for }i=1,\dots,n+1.}
\end{alignedat}
$

\vspace{1ex}

\rule{1\columnwidth}{1pt}

\caption{\label{fig:easy-strategy}Dynamic and viable execution strategy $\sigma$
for $\Gamma_{\Phi}$, when $\Phi\equiv\text{true}$. Fixed a winning
strategy $f$ for $\Phi$, the execution strategy $\sigma$ is defined
above, where $h_{i}(s)\protect\coloneqq f(s(y_{1}),\dots,s(y_{i-1}))$,
$b(s)\protect\coloneqq\min\{i\mid s(x_{i})\protect\neq h_{i}(s)\}\cup\{n+1\}$
and $b'(s)\protect\coloneqq\min\{b(s),n\}$.}

\vspace*{-1em}

\end{figure}

Notice that the value $[\sigma(s)]_{N}$ for a task $N\in{\cal T}$
and a scenario $s$ depends on the value $s(p)$ only for those variables
$p\in{\cal P}$ that are observed strictly before the time point $[\sigma(s)]_{N}$.
This condition is sufficient to guarantee that $\sigma$ is dynamic.
In particular, observe that the condition $i\leq b(s)$ depends only
on the variables $x_{1},y_{1},\dots,x_{i-1},y_{i-1}$, and that $[\sigma(s)]_{D_{i}}$
for $i\leq b'(s)$ depends on either $c_{i}^{0}$ or $c_{i}^{1}$,
whichever has been actually observed at time $(n+4)i$ in the scenario
$s$. 

One can easily check that $\sigma$ is viable, by checking that all
the constraints in ${\cal C}$ are satisfied. In particular, concerning
the constraints $(B_{n+1}\geq A_{n+1}+(n+1),\neg l_{j,1}\land\neg l_{j,2}\land\neg l_{j,3})$
for $j\in\{1,\dots,m\}$, there are two possibilities. If $b(s)=n+1$,
then all these constraint are void since each clause $(l_{j,1}\lor l_{j,2}\lor l_{j,3})$
of $\varphi$ is satisfied by the interpretation given by $s$. Otherwise,
if $b(s)\in\{1,\dots,n\}$, then they are all satisfied since $[\sigma(s)]_{B_{n+1}}=\infty\geq[\sigma(s)]_{A_{n+1}}+(n+1)=(n+4)(n+1)+n+1$.
\end{IEEEproof}

To prove the converse of Lemma~\ref{lem:easyPSPACE-hard}, we first
spot out three facts detailing out how the gadgets work as
intended. First, if the $i$-the gadget is activated,
then the controller is forced to execute either $C_{i}^{0}$ or $C_{i}^{1}$
early (Lemma~\ref{lem:hard-c0-or-c1}). Second, the activation of the $i$-th gadget and the choice of either $C_{i}^{0}$
or $C_{i}^{1}$ cannot depend on the variables $x_{i},y_{i},\dots,x_{n},y_{n}$,
so the nature can choose their values ``later on'' (Lemma~\ref{lem:hard-tech2}).
Third, if the nature copies the value selected by the
controller, the activation constraint is propagated to the next gadget
(Lemma~\ref{lem:hard-propagation}).
\begin{lem}[The controller has to schedule $C_{i}^{0}$ or $C_{i}^{1}$ early]
\label{lem:hard-c0-or-c1} Let $\sigma$ be a viable and dynamic
execution for $\Gamma_{\Phi}$. Let $s$ be any scenario and $i\in\{1,\dots,n\}$,
and suppose that $[\sigma(s)]_{B_{i}}-[\sigma(s)]_{A_{i}}\leq n$.
Then, for some $h\in\{0,1\}$, we have $[\sigma(s)]_{C_{i}^{h}}\leq[\sigma(s)]_{B_{i}}+1$.\end{lem}
\begin{IEEEproof}
Fix a scenario $s$, let $t\coloneqq[\sigma(s)]_{B_{i}}+1$, and suppose
by contradiction $[\sigma(s)]_{C_{i}^{h}}\geq t$ for $h=0,1$. Let
$d=1$ if $[\sigma(s)]_{D_{i}}\geq t$ and $d=0$ otherwise, and take
$s'=s[d/c_{i}^{0}][d/c_{i}^{1}]$. Since $t\leq[\sigma(s)]_{C_{i}^{h}}$
for $h=0,1$, we can apply Lemma~\ref{lem:dynamic-tech} obtaining
that $[\sigma(s')]_{B_{i}}+1=t$, $[\sigma(s')]_{C_{i}^{h}}\geq t$,
$[\sigma(s')]_{D_{i}}\geq t\iff[\sigma(s)]_{D_{i}}\geq t\iff d=0$,
and either $[\sigma(s)]_{A_{i}}=[\sigma(s')]_{A_{i}}<t$ or both $[\sigma(s)]_{A_{i}},[\sigma(s')]_{A_{i}}\geq t$.
Now, if $d=1$, then the constraint $D_{i}\leq B_{i}+1$ applies in
scenario $s'$ and is violated by $\sigma(s')$. Otherwise, if $d=0$,
the constraint $D_{i}\geq A_{i}+n+1$ applies in scenario $s'$ and
is violated by $\sigma(s')$. In either case, this contradicts the
fact that $\sigma$ is viable.\end{IEEEproof}
\begin{lem}[The nature can choose future variables]
\label{lem:hard-tech2} Let $\sigma$ be a viable and dynamic execution
for $\Gamma_{\Phi}$. Let $s$ be any scenario and $i\in\{1,\dots,n\}$.
Let $s'=s[v/p]$ be a scenario obtained by changing the value of any
variable $p\in\{x_{i},y_{i},\dots,x_{n},y_{n}\}$ to any value $v\in\{0,1\}$.
Then, we have $[\sigma(s')]_{B_{i}}-[\sigma(s')]_{A_{i}}\leq i-1\iff[\sigma(s)]_{B_{i}}-[\sigma(s)]_{A_{i}}\leq i-1$.
Moreover, if $[\sigma(s)]_{B_{i}}-[\sigma(s)]_{A_{i}}\leq i-1$ holds,
then $[\sigma(s')]_{C_{i}^{h}}<[\sigma(s')]_{B_{i}}+1\iff[\sigma(s)]_{C_{i}^{h}}<[\sigma(s)]_{B_{i}}+1$
for both $h=0$ and $h=1$.\end{lem}
\begin{IEEEproof}
Let $t=[\sigma(s)]_{A_{i}}+n+1$. Since $\sigma$ is viable, the unlabeled constraints
$X_{i}\geq A_{i}+(n+2)$, $Y_{i}\geq X_{i}+1$ and $A_{i+1}\geq Y_{i}+1$
imply that $t=[\sigma(s)]_{A_{i}}+n+1\leq[\sigma(s)]_{X_{i}}\leq[\sigma(s)]_{Y_{i}}\leq\dots\leq[\sigma(s)]_{X_{n}}\leq[\sigma(s)]_{Y_{n}}$.
Since $[\sigma(s)]_{A_{i}}\leq[\sigma(s)]_{A_{i}}+i-1\leq t$, Lemma~\ref{lem:dynamic-tech} can be applied to obtain that $[\sigma(s')]_{A_{i}}=[\sigma(s)]_{A_{i}}$
and $[\sigma(s)]_{B_{i}}\leq[\sigma(s)]_{A_{i}}+i-1\iff[\sigma(s')]_{B_{i}}\leq[\sigma(s')]_{A_{i}}+i-1$
which proves the first part of the lemma. Now assume $[\sigma(s')]_{B_{i}}\leq[\sigma(s)]_{A_{i}}+i-1$:
we have 
\begin{eqnarray*}
[\sigma(s')]_{B_{i}}\leq[\sigma(s')]_{B_{i}}+1 & \leq & [\sigma(s')]_{A_{i}}+(i-1)+1\\
 & \leq & [\sigma(s')]_{A_{i}}+n+1=t
\end{eqnarray*}
so by Lemma~\ref{lem:dynamic-tech} we obtain that $[\sigma(s')]_{B_{i}}=[\sigma(s)]_{B_{i}}$
and $[\sigma(s')]_{C_{i}^{h}}<[\sigma(s')]_{B_{i}}+1\iff[\sigma(s)]_{C_{i}^{h}}<[\sigma(s)]_{B_{i}}+1$.\end{IEEEproof}
\begin{lem}[Propagation of the activation constraint]
\label{lem:hard-propagation} Let $\sigma$ be a viable and dynamic
execution for $\Gamma_{\Phi}$. Let $s$ be any scenario and $i\in\{1,\dots,n\}$,
and suppose that $[\sigma(s)]_{B_{i}}-[\sigma(s)]_{A_{i}}\leq i-1$.
Let $h=0$ if $[\sigma(s)]_{C_{i}^{0}}<[\sigma(s)]_{B_{i}}+1$ and
$h=1$ otherwise. For $s'=s[h/x_{i}]$ we have $[\sigma(s')]_{B_{i+1}}-[\sigma(s')]_{A_{i+1}}\leq i$.\end{lem}
\begin{IEEEproof}
If $h=1$ then $[\sigma(s)]_{C_{i}^{1}}<[\sigma(s)]_{B_{i}}+1$
by Lemma~\ref{lem:hard-c0-or-c1}. From Lemma~\ref{lem:hard-tech2} we obtain that $[\sigma(s')]_{B_{i}}-[\sigma(s')]_{A_{i}}\leq i-1$
and $[\sigma(s')]_{C_{i}^{h}}<[\sigma(s')]_{B_{i}}+1$. Moreover,
since $\sigma$ is viable, thanks to the constraint $B_{i+1}\leq C_{i}^{h}+n+4$,
labeled $x_{i}$ if $h=1$ and $\neg x_{i}$ otherwise, we obtain
that
$
[\sigma(s)]_{B_{i+1}}  \leq  [\sigma(s)]_{C_{i}^{h}}+n+4
  <  [\sigma(s)]_{B_{i}}+1+n+4
  \leq  [\sigma(s)]_{A_{i}}+i-1+1+n+4
  \leq  [\sigma(s)]_{A_{i+1}}+i
$
where the last inequality follows from the unlabeled constraints $A_{i+1}\geq Y_{i}+1$,
$Y_{i}\geq X_{i}+1$ and $X_{i}\geq A_{i}+(n+2)$.\end{IEEEproof}
\begin{lem}
\label{lem:hardPSPACE-hard} If $\Phi\equiv\text{false}$ then $\Gamma_{\Phi}$
is not dynamically controllable.\end{lem}
\begin{IEEEproof}
Let $f\colon\{0,1\}^{*}\to\{0,1\}$ be the winning strategy of the
universal player for $\Phi$. Suppose by contradiction that $\sigma$
is a viable and dynamic execution strategy for $\Gamma_{\Phi}$.

We first construct, for $I=0,\dots,n$, step by step, a scenario
$s_{I}$ such that
\begin{enumerate}
\item[(a)]  $[\sigma(s_{I})]_{B_{I+1}}-[\sigma(s_{I})]_{A_{I+1}}\leq I$ (activation
constraint), \emph{and}
\item[(b)]  $s_{I}(y_{i})=f(s(x_{1}),\dots,s(x_{i}))$ for $i=1,\dots,I$.
\end{enumerate}
Start with any scenario $s_{0}$. We have (a) $[\sigma(s_{0})]_{B_{1}}-[\sigma(s_{I})]_{A_{1}}\leq0$
thanks to the constraint $B_{1}-A_{1}\leq0$, and there is nothing
to prove for (b). For $I=0,\dots,n-1$, define $s_{I+1}$ as follows.
Let $h_{I+1}=0$ if $[\sigma(s_{I})]_{C_{I+1}^{0}}<[\sigma(s_{I})]_{B_{I+1}}+1$
and $h_{I+1}=1$ otherwise, and define $s_{I+1}=s_{I}[h_{I+1}/x_{I+1}][f(s_{I}(x_{1}),\dots,s_{I}(x_{I}),h_{I+1})/y_{I+1}]$.
By construction (b) is satisfied. We obtain (a) by applying Lemma~\ref{lem:hard-tech2}
and Lemma~\ref{lem:hard-propagation}.

Consider the scenario $s_{n}$. We have (a) $[\sigma(s_{n})]_{B_{n+1}}-[\sigma(s_{n})]_{A_{n+1}}\leq n$.
Moreover, by (b) and the fact that $f$ is a winning strategy for
the universal player, the formula $\varphi$ is false in the interpretation
given by the scenario $s_{n}$. In particular, some clause is not
satisfied, say, the $j$-th clause for some $j\in\{1,\dots m\}$.
So, the constraint $B_{n+1}-A_{n+1}\geq n+1$ labeled with $\ell_{j}$
applies in scenario $s_{n}$, but it is violated since we proved $[\sigma(s_{n})]_{B_{n+1}}-[\sigma(s_{n})]_{A_{n+1}}\leq n$.\end{IEEEproof}
\begin{thm}
CSTN\nobreakdash-DC is PSPACE-hard.\end{thm}
\begin{IEEEproof}
Given a Q3SAT formula $\Phi$, the CSTN $\Gamma_{\Phi}$ can be easily constructed
within logarithmic internal memory. By Lemmas~\ref{lem:easyPSPACE-hard}
and~\ref{lem:hardPSPACE-hard}, it is dynamically controllable
iff $\Phi\equiv\text{true}$.
\end{IEEEproof}

\section{Polynomial-space algorithm}

\label{sec:algorithms}

\subsection{Relative execution strategies}

First, we extend some of the notions for CSTNs to the case when some
of the tasks have already been performed. This will be crucial to
describe our inductive polynomial-space algorithm.
\begin{defn}[Partial schedule, next action, completion]
A \emph{partial schedule} over ${\cal T}$ up to time $t\in\mathbb{R}$
is a schedule $\psi$ over a subset $\Dom(\psi)\subseteq{\cal T}$,
such that $[\psi]_{X}\leq t$ for every $X\in\Dom(\psi)$. Given a
partial schedule $\psi$ up to time $t$, a \emph{next action} for
$\psi$ is a pair $(t_{\mathit{next}},{\cal T}_{\mathit{next}})$
where $t_{\mathit{next}}>t$ is a time point and ${\cal T}_{\mathit{next}}\subseteq{\cal T}\setminus\Dom(\psi)$
is a non-empty set of temporal variables not assigned by $\psi$.
Let $\psi[t_{\mathit{next}}/{\cal T}_{\mathit{next}}]=\psi\cup\{(X,t_{\mathit{next}})\mid X\in{\cal T}_{\mathit{next}}\}$
be the partial schedule, up to time $t_{\mathit{next}}$, obtained
from $\psi$ by further executing all the actions in ${\cal T}_{\mathit{next}}$
at time $t_{\mathit{next}}$. Given a partial schedule $\psi$ up
to time $t$, a \emph{completion} of $\psi$ is a schedule $\psi'\in\Psi_{{\cal T}}$
such that $[\psi']_{X}=[\psi]_{X}$ for every $X\in\Dom(\psi)$ and
$[\psi']_{X}>t$ for every $X\in\Dom(\psi')\setminus\Dom(\psi)$.
Let $\Psi_{{\cal T}}[\psi]$ be the set of completions of $\psi$.
\end{defn}

\begin{defn}[Observation, completion of a partial scenario]
Given a partial scenario $h$ and a set of propositional variables
${\cal P}'\subseteq{\cal P}\setminus\Dom(h)$ not assigned by $h$,
an \emph{observation} of ${\cal P}'$ is a function $o\colon{\cal P}'\to\{0,1\}$,
and $h\cup o$ is the partial scenario obtained from $h$ by adding
all the assignments given by $o$. Given a partial scenario $h$,
a \emph{completion} of $h$ is a total scenario $s\in\Sigma_{{\cal P}}$
such that $s(p)=h(p)$ for every $p\in\Dom(h)$. Let $\Sigma_{{\cal P}}[h]$
denote the set of completions of $h$.
\end{defn}

\begin{defn}[Configuration, initial, terminal]
A \emph{configuration} is a tuple $c=(t,\psi,h)$ consisting of a time
point $t\in\mathbb{R}\cup\{-\infty\}$, a partial schedule $\psi$
up to time $t$, and a partial scenario $h\colon{\cal P}_{c}\to\{0,1\}$
where ${\cal P}_{c}=\{p\in{\cal P}\mid{\cal O}(p)\in\Dom(\psi)\}$
is the set of propositional variables observed before or at time $t$.
Let $c_{0}=(-\infty,\psi_{0},h_{0})$ be the \emph{initial configuration},
where $\Dom(\psi_{0})=\emptyset$ and $\Dom(h_{0})=\emptyset$. A
configuration $c=(t,\psi,h)$ is \emph{terminal} if, for every scenario
$s\in\Sigma_{{\cal P}}[h]$, we have ${\cal T}_{s}=\Dom(\psi)$.
\end{defn}

\begin{defn}[Next configuration]
Given a configuration $c=(t,\psi,h)$, a next action $(t_{\mathit{next}},{\cal T}_{\mathit{next}})$
for $\psi$, and an observation $o\colon{\cal P}_{\mathit{next}}\to\{0,1\}$
of ${\cal P}_{\mathit{next}}\coloneqq\{p\in{\cal P}\mid{\cal O}(p)\in{\cal T}_{\mathit{next}}\}$,
define the \emph{next configuration} $c[t_{\mathit{next}}/{\cal T}_{\mathit{next}},o]=(t_{\mathit{next}},\psi[t_{\mathit{next}}/{\cal T}_{\mathit{next}}],h\cup o)$.
\end{defn}

\begin{defn}[Relative execution strategies]
A \emph{relative execution strategy} from a configuration $c=(t,\psi,h)$
is a function $\sigma\colon\Sigma_{{\cal P}}[h]\to\Psi_{{\cal T}}[\psi]$
that maps each scenario $s$ which is a completion of $h$ to a total schedule
$\sigma(s)$, over ${\cal T}_{s}$, which is a completion of $\psi$.
A relative execution strategy $\sigma$ is \emph{viable} if $\sigma(s)$
is feasible for $\Gamma_{s}$ for every scenario $s\in\Sigma_{{\cal P}}[h]$.
It is \emph{dynamic} if, for any scenarios $s,s'\in\Sigma_{{\cal P}}[h]$
and time variable $X\in{\cal T}_{s}$, letting $t=[\sigma(s)]_{X}$,
if $\Hist(t,s,\sigma)=\Hist(t,s',\sigma)$ then $X\in{\cal T}_{s'}$
and $[\sigma(s')]_{X}=t$. Observe that a (dynamic, viable) relative
execution strategy from the initial configuration $c_{0}$ is a (dynamic,
viable) execution strategy and vice-versa.
\end{defn}

\begin{defn}[Dynamic controllability from a configuration]
A CSTN is \emph{dynamically controllable} from a configuration $c$
if it admits a dynamic and viable relative execution strategy from
$c$.
\end{defn}
The following definition and lemma serve to ensure that, in a dynamic
relative execution strategy from a non-terminal configuration, there
is always a set of actions that is executed next, all at the same
time across all the scenarios.
\begin{defn}[Well-defined next action]
A relative execution strategy $\sigma$ from a configuration $c=(t,\psi,h)$
has a \emph{well-defined next action} if there exists a next action $(t_{\mathit{next}}(\sigma), {\cal T}_{\mathit{next}}(\sigma))$ for $\psi$
such that, for every scenario $s\in\Sigma_{{\cal P}}[h]$, $\min_{X\in{\cal T}_{s}\setminus\Dom(\psi)}[\sigma(s)]_{X}=t_{\mathit{next}}(\sigma)$
and $\{X\in{\cal T}_{s}\mid[\sigma(s)]_{X}=t_{\mathit{next}}(\sigma)\}={\cal T}_{\mathit{next}}(\sigma)$.
Equivalently, $\sigma$ has a well-defined next action if, for every scenario $s\in\Sigma_{{\cal P}}[h]$, $\sigma(s)$
is a completion of $\psi[t_{\mathit{next}}(\sigma)/{\cal T}_{\mathit{next}}(\sigma)]$.
\end{defn}
\begin{lem}
\label{lem:dynamic-to-tree}If $\sigma$ is a dynamic execution strategy
from a non-terminal configuration $c=(t,\psi,h)$, then $\sigma$
has a well-defined next action.\end{lem}
\begin{IEEEproof}
Since $\sigma$ is non-terminal there exist some $s\in\Sigma_{{\cal P}}[h]$
and $X\in{\cal T}_{s}\setminus\Dom(\psi)$. Therefore, we can define
$t_{\mathit{next}}(\sigma)=\min_{s\in\Sigma_{{\cal P}}[h],X\in{\cal T}_{s}\setminus\Dom(\psi)}[\sigma(s)]_{X}$.
Then, we choose any $s_{0}\in\Sigma_{{\cal P}}[h]$ and define ${\cal T}_{\mathit{next}}(\sigma)=\{X\in{\cal T}_{s_{0}}\mid[\sigma(s_{0})]_{X}=t_{\mathit{next}}(\sigma)\}$.
We prove that ${\cal T}_{\mathit{next}}(\sigma)$ does not depend
on the choice of $s_{0}$, using the fact that $\sigma$ is dynamic,
and this concludes the proof.

Take any $s,s'\in\Sigma_{{\cal P}}[h]$, and suppose $[\sigma(s)]_{X}=t_{\mathit{next}}(\sigma)$.
We want to prove that also $[\sigma(s')]_{X}=t_{\mathit{next}}(\sigma)$.
By definition of $t_{\mathit{next}}(\sigma)$, there is no $X'\in{\cal T}\setminus\Dom(\psi)$
with either $[\sigma(s)]_{X}<t_{\mathit{next}}(\sigma)$ or $[\sigma(s')]_{X}<t_{\mathit{next}}(\sigma)$.
On the other hand, for every $X'\in\Dom(\psi)$, we have both $[\sigma(s)]_{X}<t_{\mathit{next}}(\sigma)$
and $[\sigma(s')]_{X}<t_{\mathit{next}}(\sigma)$. Hence, $\Hist(t_{\mathit{next}}(\sigma),s,\sigma)=\Hist(t_{\mathit{next}}(\sigma),s',\sigma)=h$.
Since $\sigma$ is dynamic, by applying the definition we obtain $[\sigma(s')]_{X}=t_{\mathit{next}}(\sigma)$
as desired.\end{IEEEproof}
\begin{defn}[Child configurations and strategies]
Suppose $\sigma$ has a well-defined next action. Let ${\cal P}_{\mathit{next}}(\sigma)=\{p\in{\cal P}\mid{\cal O}(p)\in{\cal T}_{\mathit{next}}(\sigma)\}$
be the set of propositional variables observed at time $t_{\mathit{next}}(\sigma)$
and $o\colon{\cal P}_{\mathit{next}}(\sigma)\to\{0,1\}$ be any outcome
for the observations. Define the \emph{child configuration} $\mathit{next}(\sigma,o)=c[t_{\mathit{next}}(\sigma)/{\cal T}_{\mathit{next}}(\sigma),o]=(t_{\mathit{next}}(\sigma),\psi[t_{\mathit{next}}(\sigma)/{\cal T}_{\mathit{next}}(\sigma)],h\cup o)$.
Since, for every $s\in\Sigma_{{\cal P}}[h]$, the schedule $\sigma(s)$
is a completion of $\psi[t_{\mathit{next}}(\sigma)/{\cal T}_{\mathit{next}}(\sigma)]$,
a relative execution strategy from $\mathit{next}(\sigma,o)$ is obtained simply restricting $\sigma$ to the scenarios that are completions of
$h\cup o$.
Denote this strategy with $\mathit{child}(\sigma,o)\coloneqq\sigma|_{\Sigma_{{\cal P}}[h\cup o]}$.\end{defn}
\begin{lem}
\label{lem:child-dynamic}If $\sigma$ is dynamic then also $\mathit{child}(\sigma,o)$
is dynamic.\end{lem}
\begin{IEEEproof}
Since $\mathit{child}(\sigma,o)$ is a restriction of $\sigma$, there
are less pairs $s,s'\in\Sigma_{{\cal P}}$ that need to be checked
in order for $\mathit{child}(\sigma,o)$ to be dynamic.\end{IEEEproof}
\begin{lem}
\label{lem:tree-to-dynamic}Let $\sigma$ be a relative execution
strategy from a non-terminal configuration $c=(t,\psi,h)$. If
$\sigma$ has a well-defined next action and
$\mathit{child}(\sigma,o)$ is dynamic for every $o\colon{\cal P}_{\mathit{next}}(\sigma)\to\{0,1\}$,
then $\sigma$ is dynamic.\end{lem}
\begin{IEEEproof}
Let $s,s'\in\Sigma_{{\cal P}}[h]$, $X\in{\cal T}_{s}$, with $t=[\sigma(s)]_{X}$
and $\Hist(t,s,\sigma)=\Hist(t,s',\sigma)$. We need to prove that
$[\sigma(s')]_{X}=t$.
If $t<t_{\mathit{next}}(\sigma)$, then $X\in\Dom(\psi)$ so $[\sigma(s)]_{X}=[\psi]_{X}=[\sigma(s')]_{X}=t$.
If $t=t_{\mathit{next}}(\sigma)$, then $X\in{\cal T}_{\mathit{next}}(\sigma)$
so $[\sigma(s')]_{X}=t_{\mathit{next}}(\sigma)=[\sigma(s)]_{X}=t$.
If $t>t_{\mathit{next}}(\sigma)$, then for any $p\in{\cal P}_{\mathit{next}}(\sigma)$,
we have $[\sigma(s)]_{{\cal O}(p)}=t_{\mathit{next}}(\sigma)<t$.
So, $p\in\Dom(\Hist(t,s,\sigma))$ and, since $\Hist(t,s,\sigma)=\Hist(t,s',\sigma)$
by hypothesis, we have $s(p)=s'(p)$. Take $o\colon{\cal P}_{\mathit{next}}(\sigma)\to\{0,1\}$
so that $o(p)=s(p)=s'(p)$ for every $p\in{\cal P}_{\mathit{next}}(\sigma)$.
 Since $\mathit{child}(\sigma,o)$ is dynamic by hypothesis, and both
$s,s'\in\Sigma_{{\cal P}}[h\cup o]$, by the definition
of dynamic strategy we get $[\sigma(s')]_{X}=t$.\end{IEEEproof}
\begin{lem}
\label{lem:child-viable}Let $\sigma$ be a dynamic relative execution
strategy from a non-terminal configuration $c$. Then, $\sigma$ is
viable iff $\mathit{child}(\sigma,o)$ is viable for every $o\colon{\cal P}_{\mathit{next}}(\sigma)\to\{0,1\}$.\end{lem}
\begin{IEEEproof}
$\sigma=\bigcup_{o\colon{\cal P}_{\mathit{next}}(\sigma)\to\{0,1\}}\mathit{child}(\sigma,o)$.\end{IEEEproof}
\begin{lem}
\label{lem:algorithm-base}A CSTN $\Gamma$ is dynamically controllable
from a \emph{terminal} configuration $c=(t,\psi,h)$ iff, for every
scenario $s\in\Sigma_{{\cal P}}[h]$, the schedule $\psi$ is feasible
for $\Gamma_{s}$.\end{lem}
\begin{IEEEproof}
There exists only one execution strategy $\sigma$ from $c$, defined
by $\sigma(s)=\psi$ for every $s\in\Sigma_{{\cal P}}[h]$. It is
clearly dynamic, and, by definition, it is viable iff, for every scenario
$s\in\Sigma_{{\cal P}}[h]$, the schedule $\psi$ is feasible for
$\Gamma_{s}$\end{IEEEproof}
\begin{lem}
\label{lem:algorithm-induction}A CSTN $\Gamma$ is dynamically controllable
from a \emph{non-terminal} configuration $c=(t,\psi,h)$ iff there
exist a next action $({\cal T}_{\mathit{next}},t_{\mathit{next}})$
from $\psi$ such that, for every observation $o\colon{\cal P}_{\mathit{next}}\to\{0,1\}$
(where ${\cal P}_{\mathit{next}}=\{p\in{\cal P}\mid{\cal O}(p)\in{\cal T}_{\mathit{next}}\}$),
$\Gamma$ is dynamically controllable from $c[t_{\mathit{next}}/{\cal T}_{\mathit{next}},o]$.\end{lem}
\begin{IEEEproof}
($\implies$) Let $\sigma$ be a dynamic and viable execution strategy
from $c$. It is sufficient to apply Lemma~\ref{lem:dynamic-to-tree}
and take $t_{\mathit{next}}=t_{\mathit{next}}(\sigma)$ and ${\cal T}_{\mathit{next}}={\cal T}_{\mathit{next}}(\sigma)$.
Then, for every $o\colon{\cal P}_{\mathit{next}}\to\{0,1\}$, the
strategy $\mathit{child}(\sigma,o)$ is dynamic and viable from $c[t_{\mathit{next}}/{\cal T}_{\mathit{next}},o]$,
thanks to Lemma~\ref{lem:child-dynamic} and Lemma~\ref{lem:child-viable}.

($\impliedby$) For every $o\colon{\cal P}_{\mathit{next}}\to\{0,1\}$,
let $\sigma_{o}$ be a dynamic and viable execution strategy from
$c[t_{\mathit{next}}/{\cal T}_{\mathit{next}},o]$. Then, define the
strategy $\sigma=\bigcup_{o\colon{\cal P}_{\mathit{next}}\to\{0,1\}}\sigma_{o}$
from the configuration $c$. Observe that $\sigma$ has a well-defined
next action, and in particular $t_{\mathit{next}}(\sigma)=t_{\mathit{next}}$
and ${\cal T}_{\mathit{next}}(\sigma)={\cal T}_{\mathit{next}}$.
Moreover, for every $o\colon{\cal P}_{\mathit{next}}\to\{0,1\}$,
we have $\mathit{child}(\sigma,o)=\sigma_{o}$ which is dynamic and
viable by assumption. Thanks to Lemma~\ref{lem:tree-to-dynamic}
and Lemma~\ref{lem:child-viable}, $\sigma$ is dynamic and viable.
\end{IEEEproof}
Lemma~\ref{lem:algorithm-base} (base case) and Lemma~\ref{lem:algorithm-induction}
(inductive case) suggest a recursive approach to solve CSTN\nobreakdash-DC.
In the inductive case, we need to consider all the possible choices
of $t_{\mathit{next}}$ and ${\cal T}_{\mathit{next}}$. However,
this is still not possible since $t_{\mathit{next}}$ is, a priori,
an unbounded real number. In the following we show that, under suitable
assumptions, we can choose $t_{\mathit{next}}$ among a finite set
of possibilities.

\subsection{Discrete strategies for CSTNs}

We assume to work on CSTNs whose constraint bounds are discrete, and
can be expressed with a finite number of bits in fixed-point precision.
This is stated in the following definition.
\begin{defn}[Discrete CSTN]
Let $w\in\mathbb{R}$ and $W\in\mathbb{N}$. A CSTN is $(w,W)$-discrete
if, for every labeled constraint $(Y\leq X+\delta,\ell)\in{\cal C}$,
we have $\delta=kw$ for $k\in\{-W,\dots,+W\}$. We call \emph{Discrete
CSTN-DC} the variant of CSTN-DC where the input CSTN is $(w,W)$-discrete
for some $w\in\mathbb{R}$ and $W\in\mathbb{N}$.
\end{defn}
We prove that, for discrete CSTNs, one can always restrict her attention to discrete execution strategies, whose execution times are expressible with a number of
bits at most polynomial in the size of the input. Our proof is a generalization
of an argument given in~\cite{Comin2015}.
\begin{defn}[Discrete execution strategy]
Let $\mu\in\mathbb{R}$ and $M\in\mathbb{N}$. A (relative) execution
strategy $\sigma$\emph{ }is \emph{$(\mu,M)$-discrete} if $[\sigma(s)]=k\mu$,
with $k\in\{1,\dots,M\}$, for every scenario $s\in\Sigma_{{\cal P}}$
and time variable $X\in{\cal T}_{s}$.\end{defn}
\begin{lem}[Discrete CSTNs admit discrete strategies]
\label{lem:cstn-discrete}Consider a $(w,W)$-discrete CSTN $\Gamma$.
If $\Gamma$ is dynamically controllable, then $\Gamma$ admits a
$(\mu,M)$-discrete viable dynamic execution strategy, for $\mu\coloneqq w/K$,
$M\coloneqq2\cdot K^{2}\cdot W$ and $K\coloneqq2^{|{\cal P}|}\cdot|{\cal T}|$.\end{lem}
\begin{IEEEproof}
Let $\sigma$ be a viable dynamic strategy for $\Gamma$. For each
$s\in\Sigma_{{\cal P}}$ and $X\in{\cal T}_{s}$, write $[\sigma(s)]_{X}$ as
\[
[\sigma(s)]_{X}=a_{s,X}\cdot W\cdot w+b_{s,X}\cdot w+c_{s,X}
\]
for $a_{s,X}\in\mathbb{Z}$, $b_{s,X}\in\{0,\dots,W-1\}$ and $c_{s,X}\in[0,w)$.
Let $A=\{a_{s,X},\,a_{s,X}+1\mid s\in\Sigma_{{\cal P}},X\in{\cal T}_{s}\}$
and $C=\{c_{s,X}\mid s\in\Sigma_{{\cal P}},X\in{\cal T}_{s}\}$. Then,
let $\alpha_{s,X}\in\{0,\dots,2K-1\}$ be the $0$-based rank respectively
of $a_{s,X}$ in $A$ and $\gamma_{s,X}\in\{0,\dots,K-1\}$ the $1$-based
rank of $c_{s,X}$ in $C$. Define the strategy $\sigma'$ as follows
\[
[\sigma'(s)]_{X}=\alpha_{s,X}\cdot W\cdot w+b_{s,X}\cdot w+\gamma_{s,X}\cdot w/K.
\]
By construction, $\sigma'$ is $(\mu,M)$-discrete for $\mu\coloneqq w/K$
and $M\coloneqq2\cdot K^{2}\cdot W$. We show that $\sigma'$ is viable
and dynamic, thus proving the statement. Observe that $[\sigma'(s)]_{X}<[\sigma'(s')]_{Y}\iff[\sigma(s)]_{X}<[\sigma(s')]_{Y}$,
for every $s\in\Sigma_{{\cal P}}$ and $X\in{\cal T}_{s}$.

(Viable.) Let $(Y\leq X+kw,\ell)$ be a constraint of $\Gamma$ and
$s\in\Sigma_{{\cal P}}$ a scenario such that $s\vDash\ell$. We have
$[\sigma(s)]_{Y}-[\sigma(s)]_{X}\leq kw$ by the assumption that $\sigma$
is viable. We prove that $[\sigma'(s)]_{Y}-[\sigma'(s)]_{X}\leq kw$,
distinguishing among the following cases.
\begin{enumerate}
\item Case $|a_{s,Y}-a_{s,X}|\geq2$.

Then also $|\alpha_{s,Y}-\alpha_{s,X}|\geq2$, since we added both
$\alpha_{s,Z}$ and $\alpha_{s,Z}+1$ to $A$, for $Z\in\{X,Y\}$.
Hence, $|[\sigma'(s)]_{Y}-[\sigma'(s)]_{X}|>W$ and, since $|k|\leq W$,
we have $[\sigma'(s)]_{Y}-[\sigma'(s)]_{X}\leq kw$.

\item Case $a_{s,Y}-a_{s,X}\in\{-1,0,+1\}$.

We have $a_{s,Y}-a_{s,X}=\alpha_{s,Y}-\alpha_{s,X}$.
\begin{enumerate}
\item Case $(a_{s,Y}\cdot W+b_{s,Y})-(a_{s,X}\cdot W+b_{s,X})\leq k-1$.

Then, $[\sigma'(s)]_{Y}-[\sigma'(s)]_{X}<(k-1)\cdot w+w\leq kw$.

\item Case $(a_{s,Y}\cdot W+b_{s,Y})-(a_{s,X}\cdot W+b_{s,X})=k$.

Then $c_{s,Y}\leq c_{s,X}$, so also $\gamma_{s,Y}\leq\gamma_{s,X}$
and $[\sigma'(s)]_{Y}-[\sigma'(s)]_{X}\leq kw+(\gamma_{s,Y}-\gamma_{s,X})\leq kw$.

\end{enumerate}
\end{enumerate}
(Dynamic.) The fact that an execution strategy is dynamic depends
only on the relative order (in $\mathbb{R}$) of the values of $[\sigma(s)]_{X}$,
which is preserved by our transformation
to $\sigma'$.
\end{IEEEproof}

\subsection{The algorithm}

We first adapt Lemma~\ref{lem:algorithm-induction} to relative execution
strategies.
\begin{defn}[Discrete configuration]
A configuration $c=(t,\psi,h)$ is $(\mu,M)$-discrete if $[\psi]_{X}=k\mu$,
with $k\in\{1,\dots,M\}$, for every time variable $X\in\Dom(\psi)$.\end{defn}
\begin{lem}
\label{lem:algorithm-induction-discrete}A CSTN $\Gamma$ admits a
$(\mu,M)$-discrete dynamic and viable execution strategy from a \emph{non-terminal}
$(\mu,M)$-discrete configuration $c=(t,\psi,h)$, iff there exist
a next action $({\cal T}_{\mathit{next}},t_{\mathit{next}})$ from
$\psi$, with $t_{\mathit{next}}=k_{\mathit{next}}\,\mu>t$ and $k_{\mathit{next}}\in\{1,\dots,M\}$,
such that, for every observation $o\colon{\cal P}_{\mathit{next}}\to\{0,1\}$
(where ${\cal P}_{\mathit{next}}=\{p\in{\cal P}\mid{\cal O}(p)\in{\cal T}_{\mathit{next}}\}$),
$\Gamma$ admits a $(\mu,M)$-discrete dynamic and viable execution
strategy from the $(\mu,M)$-discrete configuration $c[t_{\mathit{next}}/{\cal T}_{\mathit{next}},o]$.\end{lem}
\begin{IEEEproof}
Trivial adaptation of the proof of Lemma~\ref{lem:algorithm-induction}.
\end{IEEEproof}

The polynomial-memory algorithm is a mere application of Lemma~\ref{lem:cstn-discrete},
Lemma~\ref{lem:algorithm-base} and Lemma~\ref{lem:algorithm-induction-discrete}.
The pseudo-code is shown in Algorithm~\ref{alg:algorithm}.

\RestyleAlgo{ruled}
\begin{algorithm}
\caption{\label{alg:algorithm}\emph{Discrete CSTN-DC} in polynomial space.}
\SetKwInOut{Input}{Input }
\SetKwInOut{Returns}{Returns }

\SetKwInput{Assert}{Assert}

\SetKwFor{For}{for}{do}{}
\SetKwFor{Function}{Function}{}{}
\SetKwFor{RecFun}{Recursive Function}{}{}

\SetCommentSty{}
\SetKwComment{Comment}{$\triangleright$ }{}
\SetKw{Return}{return}

\SetKw{ForEachX}{foreach}
\SetKw{AndX}{and}

\SetAlgoVlined

\DontPrintSemicolon

\Function{DC$(\Gamma)$} {
    \Input{$\Gamma$ is a $(w,W)$-discrete CSTN}
    \Returns{\emph{true} if $\Gamma$ is dynamic controllable,\\ \emph{false} otherwise}

    \BlankLine

    $c_0 \coloneqq (0,\emptyset,\emptyset)$ \Comment*{Initial configuration}
    
    \Return{DC-From$(\Gamma,c_0)$} \;
}

\BlankLine

\RecFun{DC-From$(\Gamma, c)$}{
    \Input{$\Gamma$ is a $(w,W)$-discrete CSTN \\
        $c=(k\mu, \psi, h)$  is a $(\mu,M)$-discrete \\
        configuration, for $\mu\coloneqq w/K$, \\
        $M\coloneqq2\,K^{2}\cdot W$ and $K\coloneqq2^{|{\cal P}|}\cdot|{\cal T}|$.}
    \Returns{\emph{true} if $\Gamma$ is dynamic controllable from $c$,\\ \emph{false} otherwise}

    \BlankLine
    
    \If{Terminal-And-DC$(\Gamma, c)$}{
        \Return{true} \;
    }

    \BlankLine

    \Comment{Enumerate all the possible next actions ($\exists$-step)}
    \ForEach{${\cal T}_{\mathit{next}}\subseteq{\cal T}\setminus\Dom(\psi)$ not empty \AndX \\ \ForEachX $k_{\mathit{next}} \in \{k+1, \dots, M\}$}{
    
        $t_{\mathit{next}}\coloneqq k_{\mathit{next}}\,\mu$ \;
        $\psi'\coloneqq\psi[t_{\mathit{next}}/{\cal T}_{\mathit{next}}]$  \;
        
        ${\cal P}_{\mathit{next}}\coloneqq\{p\in{\cal P}\mid{\cal O}(p)\in{\cal T}_{\mathit{next}}\}$ \;
        
        \BlankLine
        
        \Comment{Enumerate all the possible observations ($\forall$-step)}
        $\mathit{AllChildrenAreDC} \gets \text{true}$ \;
        
        \ForEach{$o\colon{\cal P}_{\mathit{next}}\to\{0,1\}$}{
            $h' \coloneqq h \cup o$ \;
            $c' \coloneqq (t_{\mathit{next}},\psi',h')$ \;
            \If(\Comment*[f]{Recursion}){not DC-From$(\Gamma, c')$}{
                $\mathit{AllChildrenAreDC} \gets \text{false}$ \;
            }
        }
        \If{$\mathit{AllChildrenAreDC}$}{
            \Return{true} \;
        }
    }
    \Return{false} \;
}

\BlankLine

\Function{Terminal-And-DC$(\Gamma, c)$}{
    \Input{CSTN $\Gamma$, and configuration $c=(t, \psi, h)$}
    \Returns{\emph{true} if $c$ is a terminal configuration and \\ $\Gamma$ is dynamic controllable from $c$, \\ \emph{false} otherwise}
    
    \BlankLine
    
    \ForEach{$s\in\Sigma_{{\cal P}}[h]$}{
        \If{$\Dom(\psi)\neq{\cal T}_{s}$} {
            \Return{false} \Comment*{Not terminal}
        }
        \If{$\psi$ is not feasible for $\Gamma_{s}$} {
            \Return{false} \Comment*{Not DC}
        }
    }
    \Return{true} \;
}

\end{algorithm}

\begin{lem}
\label{lem:algorithm-space}Algorithm~\ref{alg:algorithm} can be
implemented using at most $O(|{\cal T}|\cdot(\log|W|+\log|{\cal T}|))$
space.\end{lem}
\begin{IEEEproof}
To implement the recursive procedure, it is sufficient to maintain
a stack of triples $({\cal T}_{1},k_{1},o_{1})\cdots({\cal T}_{l},k_{l},o_{l})$
containing the choices of ${\cal T}_{\mathit{next}}$, $k_{\mathit{next}}$
and $o$ at each level of the recursion. Indeed, the parameters $\psi$
and $h$ can be reconstructed from this stack. To represent the sequence
$k_{1},\dots,k_{l}$, we need $O(|{\cal T}|\log|M|)=O(|{\cal T}|(\log|W|+\log|{\cal T}|))$
bits, while ${\cal T}_{1},\dots,{\cal T}_{l}$ and $o_{1},\dots,o_{l}$
require only $O(|{\cal T}|\log|{\cal T}|)$ bits.
\end{IEEEproof}
As a consequence we get the following.
\begin{thm}
Discrete CSTN-DC is in PSPACE.
\end{thm}

\subsection{Extending to real-valued CSTNs}

It is possible to extend our polynomial algorithm so that it works
without making any assumption on how the input numbers are encoded.
Since Lemma~\ref{lem:cstn-discrete} does not apply, we need a different
way to limit the choice of $t_{\mathit{next}}$ to a finite set. We
verified that it is sufficient to take $t_{\mathit{next}}$ among
those linear combinations of the input numbers, having integer coefficients
with a number of bits polynomial in the size of the network. A formal
proof of this statement, as well as the extension of this positive
result to CSTNUs, is subject of future work.

\section{Conclusion}

\label{sec:conclusions}Our first result is a reduction from Q3SAT
to CSTN\nobreakdash-DC, which shows that checking the dynamic controllability
of CSTNs is PSPACE-hard. Our reduction relies on the close interplay
between labeled constrains and observation tasks, which
allows the planner to effectively impose her choice on some of the propositional
variables. This interplay seems to be the reason why CSTN-DC is difficult.
On the other hand, the topology of the network plays very little role.
Indeed, in our construction, the topology of the network is planar and extremely
simple: there is only one directed cycle (observe that, if there was no directed cycle, then it would be trivially and strongly controllable), and removing a \emph{single} edge, from $A_{1}$ to $B_{1}$,
we get an acyclic graph, and actually an inward arborescence (a directed
rooted tree, where all edges point towards the root) when disregarding
the parallel edges between $A_{n}$ and $B_{n}$.

Our second result is an algorithm for CSTN\nobreakdash-DC that uses
only polynomial space, proving that $\text{CSTN\nobreakdash-DC}\in\text{PSPACE}$.
This PSPACE algorithm actually searches for a viable dynamic execution
strategy and can be easily implemented as to return the one found. However, since our algorithm
works by brute force, guessing the execution times among a finite
but large set of possibilities, it does not seem suitable to be applied
in practice. Nevertheless, by showing that polynomial space is sufficient
to solve this problem, we open up the challenge of finding a practical
algorithm, that only requires polynomial memory in the worst case.

\bibliographystyle{IEEEtran}
\bibliography{references}

\begin{thebibliography}{10}
\providecommand{\url}[1]{#1}
\csname url@samestyle\endcsname
\providecommand{\newblock}{\relax}
\providecommand{\bibinfo}[2]{#2}
\providecommand{\BIBentrySTDinterwordspacing}{\spaceskip=0pt\relax}
\providecommand{\BIBentryALTinterwordstretchfactor}{4}
\providecommand{\BIBentryALTinterwordspacing}{\spaceskip=\fontdimen2\font plus
\BIBentryALTinterwordstretchfactor\fontdimen3\font minus
  \fontdimen4\font\relax}
\providecommand{\BIBforeignlanguage}[2]{{%
\expandafter\ifx\csname l@#1\endcsname\relax
\typeout{** WARNING: IEEEtran.bst: No hyphenation pattern has been}%
\typeout{** loaded for the language `#1'. Using the pattern for}%
\typeout{** the default language instead.}%
\else
\language=\csname l@#1\endcsname
\fi
#2}}
\providecommand{\BIBdecl}{\relax}
\BIBdecl

\bibitem{Dechter1991}
R.~Dechter, I.~Meiri, and J.~Pearl, ``{Temporal constraint networks},''
  \emph{Artificial Intelligence}, vol.~49, no. 1-3, pp. 61--95, May 1991.

\bibitem{Tsamardinos2003a}
I.~Tsamardinos, T.~Vidal, and M.~E. Pollack, ``{CTP}: A new constraint-based
  formalism for conditional, temporal planning,'' \emph{Constraints}, vol.~8,
  no.~4, pp. 365--388, 2003.

\bibitem{Vidal1999}
T.~Vidal, ``Handling contingency in temporal constraint networks: from
  consistency to controllabilities,'' \emph{Journal of Experimental and
  Theoretical Artificial Intelligence}, vol.~11, no.~1, pp. 23--45, Jan. 1999.

\bibitem{Morris2001}
P.~Morris, N.~Muscettola, and T.~Vidal, ``{Dynamic control of plans with
  temporal uncertainty},'' in \emph{IJCAI International Joint Conference on
  Artificial Intelligence}, 2001, pp. 494--499.

\bibitem{Comin2015}
C.~Comin and R.~Rizzi, ``Dynamic consistency of conditional simple temporal
  networks via mean payoff games: a singly-exponential time {DC}-checking,''
  \emph{arXiv:1505.00828 [cs]}, May 2015.

\bibitem{Cimatti2014}
A.~Cimatti, L.~Hunsberger, A.~Micheli, R.~Posenato, and M.~Roveri, ``Sound and
  complete algorithms for checking the dynamic controllability of temporal
  networks with uncertainty, disjunction and observation,'' in \emph{2014 21st
  International Symposium on Temporal Representation and Reasoning}.\hskip 1em
  plus 0.5em minus 0.4em\relax IEEE, Sep. 2014, pp. 27--36.

\bibitem{Hunsberger2015a}
L.~Hunsberger, R.~Posenato, and C.~Combi, ``{A Sound-and-Complete
  Propagation-Based Algorithm for Checking the Dynamic Consistency of
  Conditional Simple Temporal Networks},'' in \emph{2015 22nd International
  Symposium on Temporal Representation and Reasoning (TIME)}.\hskip 1em plus
  0.5em minus 0.4em\relax IEEE, Sep. 2015, pp. 4--18.

\bibitem{Hunsberger2012}
------, ``The dynamic controllability of conditional {STN}s with uncertainty,''
  pp. 1--8, 2012.

\bibitem{Combi2013}
C.~Combi, L.~Hunsberger, and R.~Posenato, ``An algorithm for checking the
  dynamic controllability of a conditional simple temporal network with
  uncertainty,'' \emph{Evaluation}, 2013.

\bibitem{Combi2014}
------, ``An algorithm for checking the dynamic controllability of a
  conditional simple temporal network with uncertainty - revisited,'' in
  \emph{International Conference on Agents}, 2014, pp. 314--331.

\end{thebibliography}

\end{document}